\documentclass{webofc}
\usepackage[varg]{txfonts}   
\begin{document}
\title{Constraints on hadron resonance gas interactions via first-principles Lattice QCD susceptibilities}
%
%

\author{\firstname{Jamie M.} \lastname{Karthein}\inst{1}\fnsep\thanks{\email{jmkar@mit.edu}} \and
        \firstname{Volker} \lastname{Koch}\inst{2} \and
        \firstname{Claudia} \lastname{Ratti}\inst{3} \and
        \firstname{Volodymyr} \lastname{Vovchenko}\inst{2,3,4}
}
\institute{Center for Theoretical Physics, Massachusetts Institute of Technology, Cambridge, MA 02139, U.S.A.
\and
           Nuclear Science Division, Lawrence Berkeley National Laboratory, Berkeley, CA 94720, U.S.A. 
\and
           Department of Physics, University of Houston, Houston, TX 77204, U.S.A.
\and
            Institute for Nuclear Theory, University of Washington, Seattle, WA 98195, U.S.A.
          }

\abstract{%
  We investigate extensions of the Hadron Resonance Gas (HRG) Model beyond the ideal case by incorporating both attractive and repulsive interactions into the model. When considering additional states exceeding those measured with high confidence by the Particle Data Group, attractive corrections to the overall pressure in the HRG model are imposed. On the other hand, we also apply excluded-volume corrections, which ensure there is no overlap of baryons by turning on repulsive (anti)baryon-(anti)baryon interactions. We emphasize the complementary nature of these two extensions and identify combinations of conserved charge susceptibilities that allow us to constrain them separately. In particular, we find interesting ratios of susceptibilities that are sensitive to one correction and not the other. This allows us to constrain the excluded volume and particle spectrum effects separately. Analysis of the available lattice results suggests the presence of both the extra states in the baryon-strangeness sector and the repulsive baryonic interaction, with indications that hyperons have a smaller repulsive core than non-strange baryons. We note that these results are interesting for heavy-ion-collision systems at both the LHC and RHIC.
}
\maketitle
\section{Introduction}
\label{intro}
The phase structure of strongly-interacting matter as mapped out in the QCD phase diagram represents highly sought after information on the rich phenomena associated with the underlying theory.
Along the temperature axis, i.e. at vanishing baryon chemical potentials, the transition from ordinary nuclear matter to the Quark-Gluon Plasma is known to be a crossover from first-principles Lattice QCD techniques \cite{Aoki:2006we}.
In addition, the low temperature phase of strongly-interacting matter is well-described by a gas of hadronic resonances \cite{Borsanyi:2010bp,Vovchenko:2016rkn}.
Its  agreement with the equation of state from lattice calculations has led to its wide applications, especially in the study of the chemical freeze-out in heavy-ion collisions (HICs) \cite{Vovchenko:2015idt,Bellwied:2018tkc,Alba:2020jir}. 
However, more differential observables like susceptibilities of conserved charges have revealed discrepancies between the predictions of the fundamental theory and the hadronic phase treatment \cite{Bazavov:2014xya,Borsanyi:2018grb}, specifically in the transition region of the QCD phase diagram.
As laid out in Ref. \cite{Karthein:2021cmb}, we consider two HRG model extensions: an extended hadronic spectrum and the excluded volume interaction in the baryon sector. 
The resulting HRG model considerably improves the description of lattice QCD results for fluctuations of conserved charges.

\section{Beyond the Ideal HRG Model}
\label{model}


In the Boltzmann approximation, the partial pressures in the ideal HRG model for particle species $i$ can be written in the following form:
\begin{equation} \label{eq:press_phi}
\centering
    p_i (T, \mu_B, \mu_Q, \mu_S) = d_i \tilde{\phi}(T,m_i) \lambda_i(T,\mu_i), \quad
    \tilde{\phi}(T,m_i) =  \frac{m_i^2 T^2}{2\pi^2} \, K_2(m_i/T).
\end{equation}
The full pressure is then given as the sum of the partial pressures of all hadronic sectors $i$ with various $BQS$ quantum numbers:
\begin{equation} \label{eq:Pfug}
    \centering
    p (T, \mu_B, \mu_Q, \mu_S) = \tilde{\phi_0}(T) + \sum_{i \neq 0} 2 \, \tilde{\phi_i}(T) \cosh \left(\mu_i/T \right),
\end{equation}
where $\mu_i=B_i \mu_B + Q_i \mu_Q + S_i \mu_S$ is the chemical potential of the corresponding $i^{th}$ sector and $\tilde{\phi_0}(T)$ is the $i=0$ term, representing charge neutral species. 
Each term in Eq.~\eqref{eq:Pfug} corresponds to the partial pressure associated with the particular set of hadronic quantum numbers.
The complementary extensions to the ideal HRG model explored in this work involve supplementing the well-known resonances from the Particle Data Booklet \cite{Zyla:2020zbs} and prohibiting the overlap of baryon states.

Our first modification to the standard HRG model is the incorporation of hadronic states beyond the experimentally well-known ones. 
The PDG provides an assessment of how well established various measured hadronic states are, with a rating based on a number of stars (*). 
The **** states are those which are very well known, such as protons or $\Delta(1232)$ resonances. 
Conversely, the * states are the least established, for example high-mass resonances like $\Delta$(1750). 
The three hadronic lists under study here are as follows: 
PDG2016 -- ordinary hadronic list (*** $-$ ****) from the 2016 Particle Data Booklet;
PDG2016+ -- hadronic list with established and unconfirmed~(*$-$****) states;
Quark Model~(QM) -- the list which incorporates all states predicted by the Quark Model.
The latter two lists were introduced and described in detail in \cite{Alba:2017mqu, Alba:2020jir}.
An update to the QM list was provided in Ref. \cite{Karthein:2021cmb} based on the findings in Ref. \cite{Bollweg:2021vqf}. 

The next extension to the HRG model is the excluded volume model. This corresponds to including repulsive interactions between hadrons. Many versions of the EV-HRG model have been considered in the literature. Here we follow the approach introduced in Refs.~\cite{Vovchenko:2016rkn,Vovchenko:2017xad} where EV interactions are included only for baryon-baryon and antibaryon-antibaryon pairs. 
This corresponds to a minimalistic EV extension that does not affect meson-meson and meson-baryon interactions, which are presumed to be dominated by resonance formation and thus already included in the HRG model.
The pressure is partitioned into contributions of non-interacting mesons and interacting baryons and antibaryons: $p = p_M^{\rm id} + p_B^{\rm ev} + p_{\bar{B}}^{\rm ev}$.
\begin{equation} \label{eq:pBev}
\centering
p_{M}^{\rm id}  = \tilde{\phi}_0(T) ~ + \sum_{i \neq 0, \, i \in M} 2 \, \tilde{\phi}_i(T) \, \cosh(\mu_i/T), \quad
p_{B(\bar{B})}^{\rm ev}  =  \sum_{i \in B} \tilde{\phi}_i(T) \, \exp(\pm \mu_i/T) \,
\exp\left( \frac{- b \, p_{B(\bar{B})}^{\rm ev}}{T} \right)
\end{equation}
Here, $i \in M$ corresponds to mesons~($B_i = 0$), $i \in B$ corresponds to baryons~($B_i = 1$), $b$ is the baryon excluded volume parameter, and $\tilde{\phi}(T)$ is given in Eq. (\ref{eq:press_phi}).
Equation~\eqref{eq:pBev} can be solved in terms of the Lambert W function:
\begin{equation}\label{eq:pBevW}
p_{B(\bar{B})}^{\rm ev}  = \frac{T}{b} \, W[\varkappa_{B(\bar{B})}(T,\mu_B,\mu_Q,\mu_S)], \quad
\varkappa_{B(\bar{B})}(T,\mu_B,\mu_Q,\mu_S) = b \, \sum_{i \in B} \tilde{\phi}_i(T) \, \exp(\pm \mu_i/T).
\end{equation}
%
%

\section{Results}
\label{results}
In order to constrain the hadron spectrum,  we consider ratios of second order susceptibilities: $ \chi_{11}^{BQ} / \chi_2^B, \chi_{11}^{BS} / \chi_2^B$,
in which the excluded volume parameter $b$ cancels out.
On the other hand, the ratios are sensitive to the particle list via the ``partial pressures'' $\tilde{\phi}_j$ from Eq. \ref{eq:Pfug}.
%
\begin{figure*}
    \centering
    \includegraphics[width=0.44\linewidth]{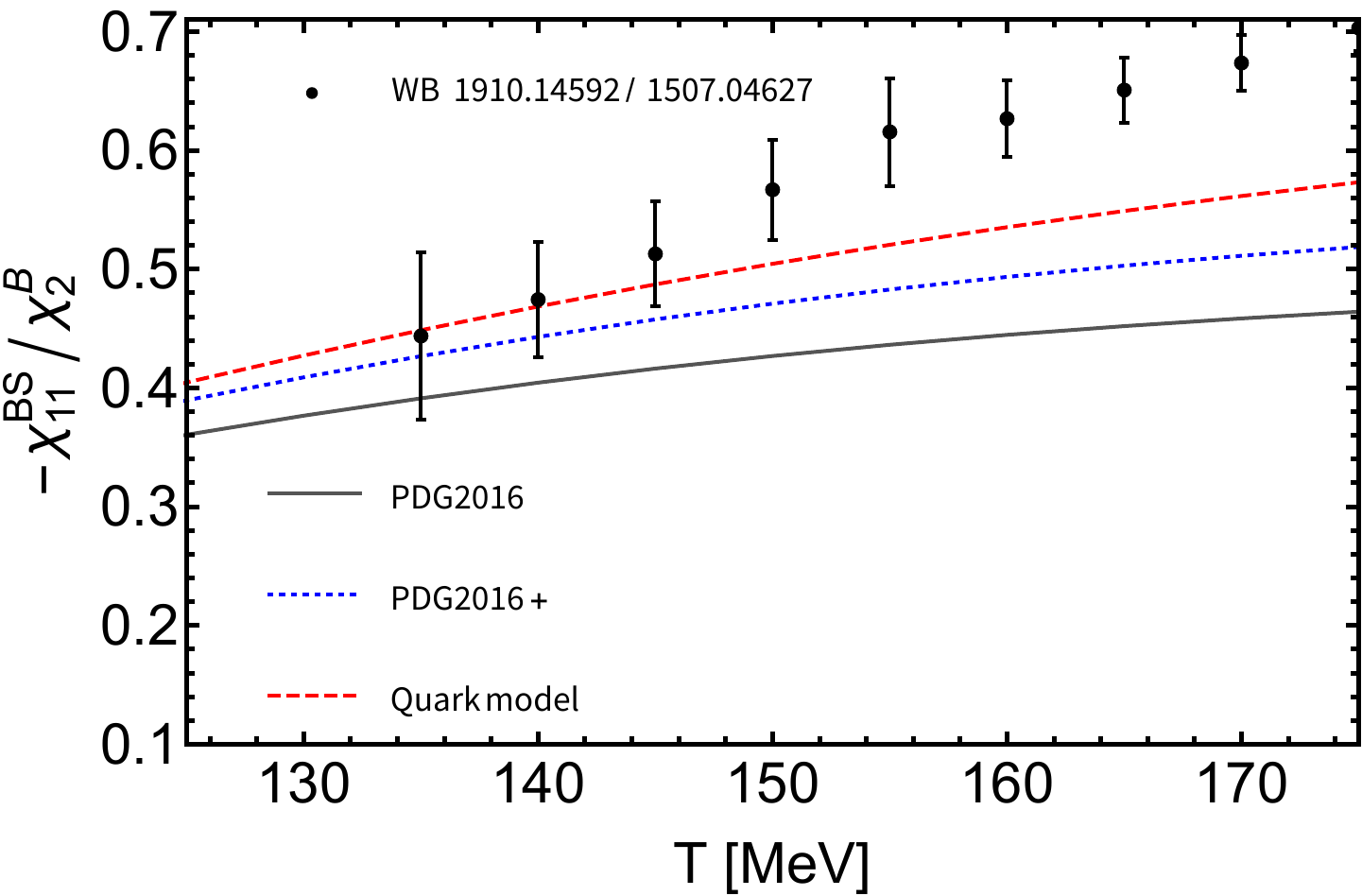}
    \includegraphics[width=0.44\linewidth]{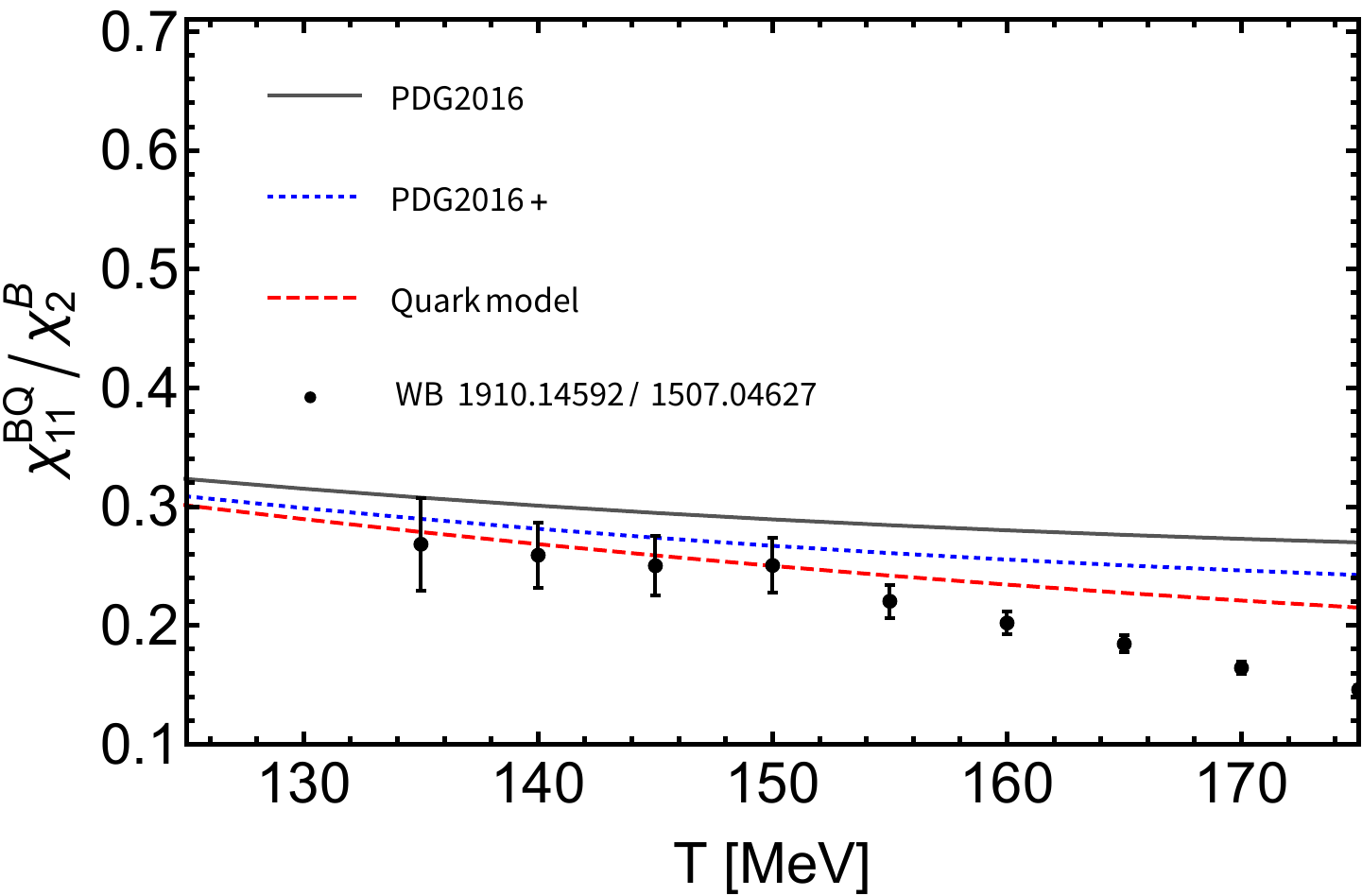}
    \includegraphics[width=0.45\linewidth]{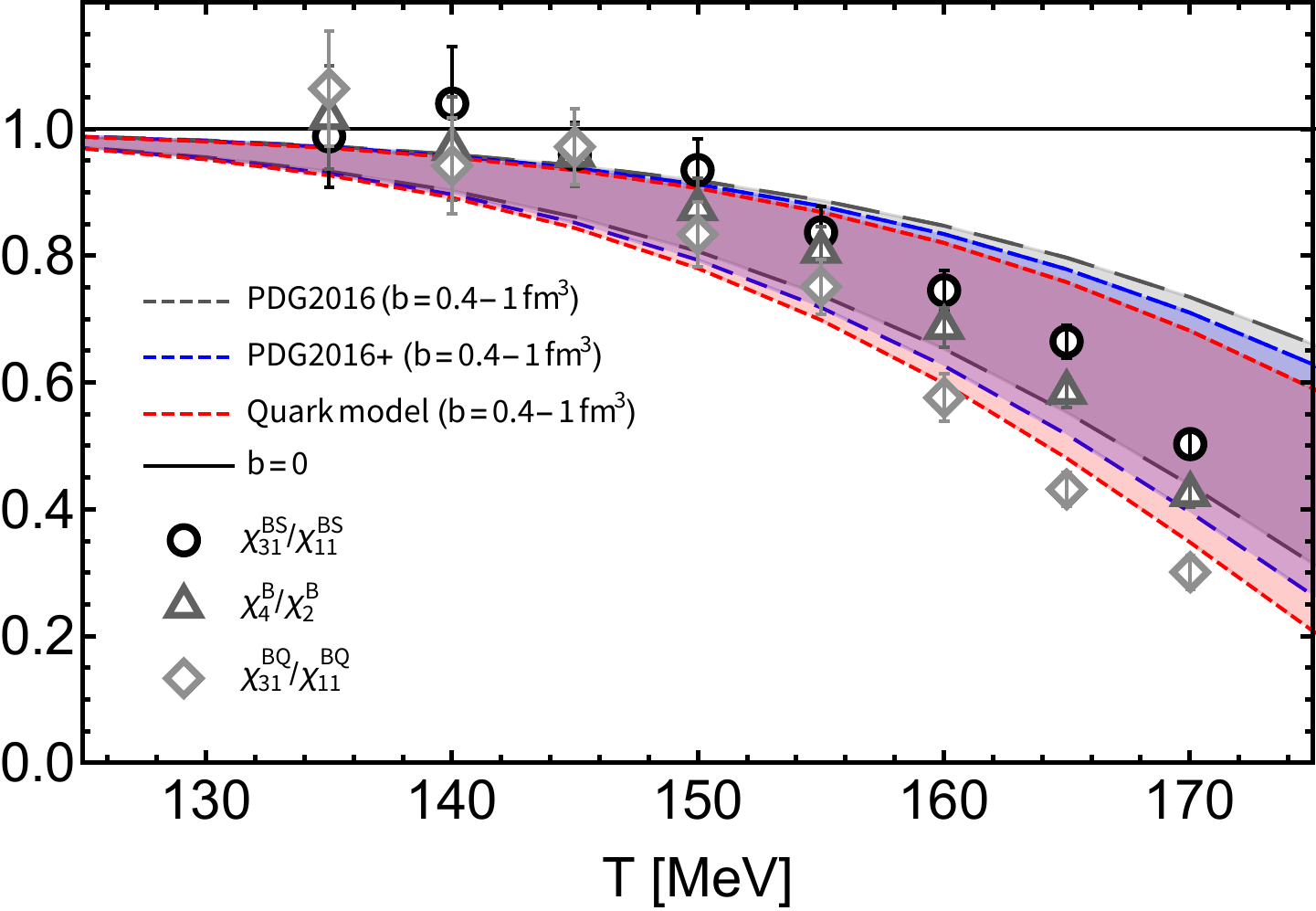}
    \caption{Temperature dependence of susceptibility ratios in the EV-HRG model (colored lines) and from Lattice QCD (grayscale points). Top: second order susceptibility ratios insensitive to the excluded volume parameter $b$. Bottom: fourth-to-second order susceptibility ratios, insensitive to the hadron spectrum. Here the colored bands correspond to a range of $b: 0.4-1$ fm.}
    \label{fig:chi-BS-2B_no_EV_dep}
\end{figure*}
%
The ratios $\chi_{11}^{BQ}/\chi_2^B$ and $\chi_{11}^{BS}/\chi_2^B$ are shown in Fig.~\ref{fig:chi-BS-2B_no_EV_dep}.
The comparison with continuum extrapolated lattice data \cite{Bellwied:2019pxh, Bellwied:2015lba} suggests the need for additional hyperon states from PDG2016+/QM, as previously discussed in Ref. \cite{Alba:2017mqu}.

Next, we can place limits on the excluded volume parameter, $b$ by considering ratios of fourth-to-second order susceptibilities.
From Eq. \ref{eq:pBevW}, $\varkappa_B = 0$ corresponds to the ideal HRG model, leading to $\frac{\chi^B_4}{\chi^B_2} = \frac{\chi^{BS}_{31}}{\chi^{BS}_{11}} = \frac{\chi^{BQ}_{31}}{\chi^{BQ}_{11}} = 1$ regardless of the inclusion of any additional hadronic states. 
Thus, these three ratios maintain their equality in the EV-HRG model under consideration and are sensitive to the EV parameter $b$.
Furthermore, given the previous equality within the model, we can estimate the limits of validity of the model as the point where the lattice results of these quantities begin to deviate from one another.
Generally speaking, the behavior of the three ratios is very similar over the temperature range, as shown in Fig. \ref{fig:chi-BS-2B_no_EV_dep}.
Quantitatively, however, we see that when $T \leq 150$ MeV the three ratios sit on top of each other. 
Where they fall out of agreement, i.e. around $T \sim 160$ MeV, could indicate the breakdown of the model, since these ratios are predicted to be identical in the EV-HRG model.

\section{Conclusions}
\label{conclusions}
We have implemented additional interactions in the Hadron Resonance Gas model with the motivation to achieve the best agreement with first-principles Lattice QCD data. 
By utilizing the lattice results on conserved charge susceptibilities, we were able to constrain the free parameters, namely the hadron spectrum and the excluded volume parameter $b$. 
We find that the hadron spectrum needs further contribution from strange resonances beyond the well-known PDG states.
On the other hand, the excluded volume is constrained to $0.4 \leq b \leq 1 \text{fm}^3$ by studying fourth-to-second order ratios.
The EV-HRG predicts these ratios to be identical, however, the lattice data reveal statistically significant differences between the three ratios at $T \gtrsim 155$~MeV, which follow a hierarchy $\chi_{31}^{BS}/\chi_{11}^{BS} > \chi_4^B/\chi_2^B > \chi_{31}^{BQ}/\chi_{11}^{BQ}$.
We argue that this hierarchy indicates a flavor dependence in the baryon excluded volumes, namely that strange baryons have generally smaller excluded volumes than non-strange baryons.

\section{Acknowledgements}
\label{acknowledgements}
This material is based upon work supported by the National Science Foundation under grant no. PHY1654219, PHY2208724 and PHY-2116686, by the U.S. Department of Energy, Office of Science, Office of Nuclear Physics, under contract number DE-AC02-05CH11231231, and within the framework of the Beam Energy Scan Theory (BEST) Collaboration. This work was supported by the National Science Foundation (NSF) within the framework of the MUSES collaboration, under grant number OAC-2103680. J.M.K. is supported by an NSF Ascending Postdoctoral Scholar Fellowship under Award No. 2138063. V.V. acknowledges support via the Feodor Lynen program of the Alexander von Humboldt foundation.


%
\bibliography{all}

\begin{thebibliography}{15}

\bibitem{Aoki:2006we}
Y.~Aoki, G.~Endrodi, Z.~Fodor, S.D. Katz, K.K. Szabo, Nature \textbf{443}, 675
  (2006), \texttt{hep-lat/0611014}

\bibitem{Borsanyi:2010bp}
S.~Borsanyi, Z.~Fodor, C.~Hoelbling, S.D. Katz, S.~Krieg, C.~Ratti, K.K. Szabo
  (Wuppertal-Budapest), JHEP \textbf{09}, 073 (2010), \texttt{1005.3508}

\bibitem{Vovchenko:2016rkn}
V.~Vovchenko, M.I. Gorenstein, H.~Stoecker, Phys. Rev. Lett. \textbf{118},
  182301 (2017), \texttt{1609.03975}

\bibitem{Vovchenko:2015idt}
V.~Vovchenko, V.V. Begun, M.I. Gorenstein, Phys. Rev. C \textbf{93}, 064906
  (2016), \texttt{1512.08025}

\bibitem{Bellwied:2018tkc}
R.~Bellwied, J.~Noronha-Hostler, P.~Parotto, I.~Portillo~Vazquez, C.~Ratti,
  J.M. Stafford, Phys. Rev. \textbf{C99}, 034912 (2019), \texttt{1805.00088}

\bibitem{Alba:2020jir}
P.~Alba, V.M. Sarti, J.~Noronha-Hostler, P.~Parotto, I.~Portillo-Vazquez,
  C.~Ratti, J.M. Stafford, Phys. Rev. C \textbf{101}, 054905 (2020),
  \texttt{2002.12395}

\bibitem{Bazavov:2014xya}
A.~Bazavov et~al., Phys. Rev. Lett. \textbf{113}, 072001 (2014),
  \texttt{1404.6511}

\bibitem{Borsanyi:2018grb}
S.~Borsanyi, Z.~Fodor, J.N. Guenther, S.K. Katz, K.K. Szabo, A.~Pasztor,
  I.~Portillo, C.~Ratti, JHEP \textbf{10}, 205 (2018), \texttt{1805.04445}

\bibitem{Karthein:2021cmb}
J.M. Karthein, V.~Koch, C.~Ratti, V.~Vovchenko, Phys. Rev. D \textbf{104},
  094009 (2021)

\bibitem{Zyla:2020zbs}
P.~Zyla et~al. (Particle Data Group), PTEP \textbf{2020}, 083C01 (2020)

\bibitem{Alba:2017mqu}
P.~Alba et~al., Phys. Rev. \textbf{D96}, 034517 (2017), \texttt{1702.01113}

\bibitem{Bollweg:2021vqf}
D.~Bollweg, J.~Goswami, O.~Kaczmarek, F.~Karsch, S.~Mukherjee, P.~Petreczky,
  C.~Schmidt, P.~Scior (HotQCD), Phys. Rev. D \textbf{104} (2021),
  \texttt{2107.10011}

\bibitem{Vovchenko:2017xad}
V.~Vovchenko, A.~Pasztor, Z.~Fodor, S.D. Katz, H.~Stoecker, Phys. Lett. B
  \textbf{775}, 71 (2017), \texttt{1708.02852}

\bibitem{Bellwied:2019pxh}
R.~Bellwied, S.~Borsanyi, Z.~Fodor, J.N. Guenther, J.~Noronha-Hostler,
  P.~Parotto, A.~Pasztor, C.~Ratti, J.M. Stafford (2019), \texttt{1910.14592}

\bibitem{Bellwied:2015lba}
R.~Bellwied, S.~Borsanyi, Z.~Fodor, S.~Katz, A.~Pasztor, C.~Ratti, K.~Szabo,
  Phys. Rev. D \textbf{92}, 114505 (2015), \texttt{1507.04627}

\end{thebibliography}
%
%

\end{document}